\documentclass[a4paper,11pt]{article}
\usepackage{amsmath}
\usepackage{amssymb}
\usepackage{epsfig}
\usepackage{graphics}

\begin{document}




\thispagestyle{empty}
\begin{titlepage}

\vspace*{-1cm}
\hfill \parbox{3.5cm}{BUTP-99/08 \\ 
BUHE-9901} 
\vspace*{1.0cm}

\begin{center}
  {\large {\bf \hspace*{-0.2cm}
Glueball production in hadron and nucleus collisions
  }
      \footnote{Work
      supported in part by the Schweizerischer Nationalfonds.}  }
  \vspace*{3.0cm} \\

{\bf

	S. Kabana} \\
	Laboratory for High Energy Physics\\

	and\\

{\bf    P. Minkowski} \\
     Institute for Theoretical Physics \\

     University of Bern \\
     Sidlerstrasse 5 \\
     CH - 3012 Bern , Switzerland
   \vspace*{0.8cm} \\

15. June 1999

\vspace*{2.0cm}

\begin{abstract}
\noindent
We 
elaborate on the hypothesis that in high energy hadron
hadron and nucleus nucleus
collisions the lowest mass glueballs
are copiously produced from the gluon rich environment 
especially at high energy density.
We discuss the particular glueball decay modes: $0^{++}, 2^{++} 
\rightarrow K \overline{K}$ and 
$0^{++} \rightarrow \pi^{+} \ \pi^{-} \ \ell^{\ +} \ \ell^{-}$.

\end{abstract}
\end{center}

\end{titlepage}







\pagestyle{plain}




\section{Introduction}

\noindent
Early ideas on properties of gluonic mesons or glueballs as bound states
of gauge bosons within QCD remained inconclusive at first because
of an apparent lack of phenomenological evidence for such resonances
beyond the prominent mesons $q \overline{q}$, baryons $qqq$ 
and antibaryons $\overline{q} \overline{q} \overline{q}$. 

\noindent
An outline of expected scenarios for (mainly) binary glueballs,
i.e. consisting predominantly of two gauge bosons, was presented in ref. \cite{HFPM}.
Specifically a low mass and high mass scenario were distinguished in ref. \cite{HFPM},
referring to the mass of the lowest lying 
glueball with quantum numbers $J^{\ PC} \ = \ 0^{++}$.

\noindent
The latter is expected to be close to 1 GeV in the low mass scenario, whereas it is to be
sought above or around 1.5 GeV in the high mass case. For an appraisal of these ideas
we refer to ref. \cite{Close}. Lattice gauge theory, in particular without quark flavors 
indicates preference for the high mass scenario \cite{Michael}, \cite{Teper}, 
\cite{latt}, \cite{latt2}.
In the presence of quarks lattice simulations are at the moment
inconclusive \cite{lattunq1}, \cite{lattunq2}.

\section{Reference production of glueballs}

\noindent
In a recent paper \cite{PMWO} the spectroscopic assignment of the
three lowest lying binary gluonic mesons denoted 
$gb \ ( \ 0^{ ++} \ ) \ , \ gb \ ( \ 0^{-+} \ ) \ , \ gb \ ( \ 2^{++} \ )$
shown in table \ref{tabsum} was proposed and discussed. 

\begin{table}[ht]
\[
\begin{array}{c@{\hspace*{0.6cm}}c@{\hspace*{0.4cm}}
l@{\hspace*{0.4cm}}c@{\hspace*{0.4cm}}c}\\ \hline
  \mbox{name} & \mbox{PDG} & \mbox{mass (MeV)}& \mbox{mass}^2 \mbox{(GeV)}^2 
& \mbox{width (MeV)}
  \\ \hline 
   gb \ ( \ 0^{++} \ ) & f_0(400-1200)& \sim 1000 & \sim 1. &
      \hspace*{0.2cm}   500-1000
  \vspace*{0.1cm}   \\
  & f_0(1370) &&&
        \vspace*{0.1cm}\\
   g b \ ( \ 0^{-+} \ ) & \eta  (1440)& 1400\ -\ 1470 & 2.07 &
      \hspace*{0.2cm} 50\ -\ 80
  \vspace*{0.1cm} \\
   gb \ ( \ 2^{++} \ ) & f_{J}  (1710)& 1712 \ \pm \ 5 & 2.93
 & 133 \ \pm \ 14 \\
\hline
\end{array}
\]
\caption{Properties of the basic triplet of binary glueballs.}
\label{tabsum}
\end{table}

\noindent
Two main modes of production of $ gb ( 0^{++} )$  were elaborated on in ref. \cite{PMWO} :

\begin{description}
\item
a) two to two pseudoscalar meson scattering in the I = 0 channel :

\begin{table}[ht]
\[
\begin{array}{c@{\hspace*{0.2cm}}c@{\hspace*{0.2cm}}c@{\hspace*{0.2cm}}c@{\hspace*{0.2cm}}
c@{\hspace*{0.2cm}}}
\pi \ \pi & \rightarrow & gb \ ( \ 0^{++} \ ) & \rightarrow & \pi \ \pi 
  \vspace*{0.1cm}   \\
          & \rightarrow & gb \ ( \ 0^{++} \ ) & \rightarrow & K \ \overline{K} 
  \vspace*{0.1cm}   \\
          & \rightarrow & gb \ ( \ 0^{++} \ ) & \rightarrow & \eta \ \eta
\end{array}
\]
\caption{Production and decay modes of gb $0^{++}$ in quasi-elastic
$\pi \pi$ scatering.}
\label{elast}
\end{table}

The wide structure interrupted by two narrow width resonances,
which manifest themselves as almost complete negative interference
is referred to as 'red dragon' in ref. \cite{PMWO}. It is interpreted as $gb \ ( \ 0^{++} \ )$
interfering with the two scalar, isoscalar $q \overline{q}$ mesons,
($f_0(980)$, $f_0(1500)$) 
forming the three flavor nonet.

This contrasts with the assignment proposed in ref. \cite{amsclo} where 
$f_{0} \ (1500)$ is identified with $gb \ ( \ 0^{++} \ )$ together with a nontrivial
admixture of scalar $q \overline{q}$ states.

The production characteristics is represented by the modulus square of the s wave
scattering amplitude in the three channels in table \ref{elast},
which we reproduce in figure \ref{gbfig3} from ref. \cite{PMWO}.

\begin{figure}[ht]
\begin{center}
\mbox{\epsfig{file=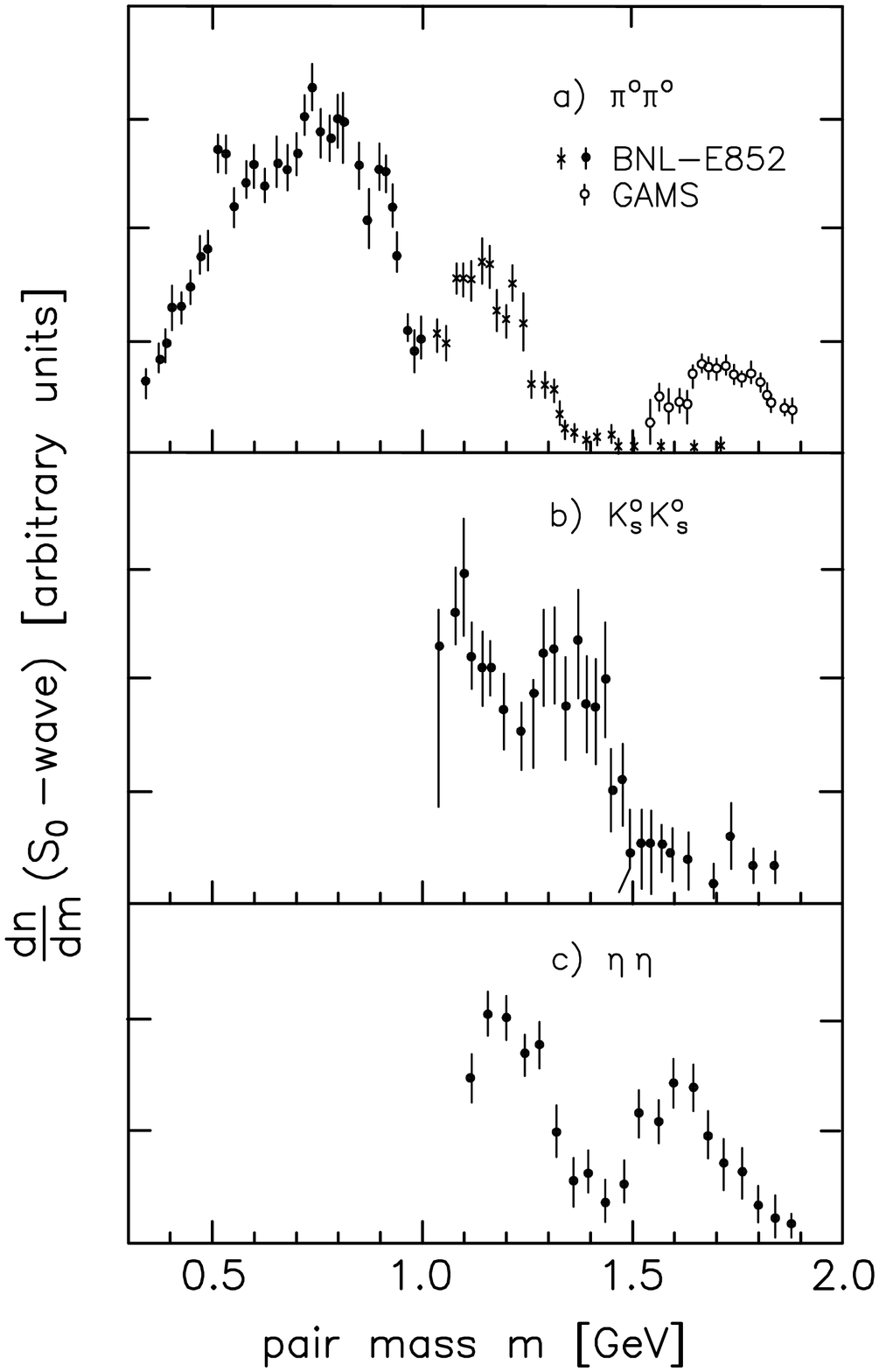,width=100mm}}
\end{center}
\caption{
Isoscalar S-wave components of the mass spectra of
pseudoscalar pairs produced in $\pi p$-collisions at small
momentum transfers t, 
(a) $\pi^0 \pi^0$ spectrum, the preferred solution for $m < 1.5$ GeV
by the BNL-E852 experiment \protect\cite{bnl_e852}
 (preliminary results) 
 and
 an alternative solution for higher masses by GAMS \protect\cite{GAMS};
 (b) $K^0_s K^0_s$ spectrum by Etkin et al. \protect\cite{etkin}
 which is similar to the results by Cohen et al. \protect\cite{cohen}
below 1600 MeV and (c) 
$\eta \eta$ spectrum by Binon et al. \protect\cite{binon}.
}
\label{gbfig3}
\end{figure}

\item
b) Production in the reaction $p \ p \ \rightarrow \ p \ X \ p$ 
at central rapidity.

A most remarkable line shape over the whole region of $gb \ ( \ 0^{++} \ )$ 
displaying a width of approximately 1 GeV is observed in the
double pomeron dominated reaction
\begin{equation}
  \label{eq:0}
p_{\ 1} \ + \ p_{\ 2} \ \rightarrow \ p'_{\ 1} \ + \ X \ + \ p'_{\ 2} 
\end{equation}
\noindent
where the momentum transfers from $p_{\ 1}$ to $p'_{\ 1}$ and from
$p_{\ 2}$ to $p'_{\ 2}$ are constrained to be quite small compared to 
the center of mass energy.

\noindent
The line shape as observed by the Axial Field Spectrometer collaboration
\cite{akesson} is shown in figure \ref{figcent}.
It  represents the production of the $0^{++}$ glueball state
(called red dragon) in ref. \cite{PMWO}.


\begin{figure}[ht]
\begin{center}
\mbox{\epsfig{file=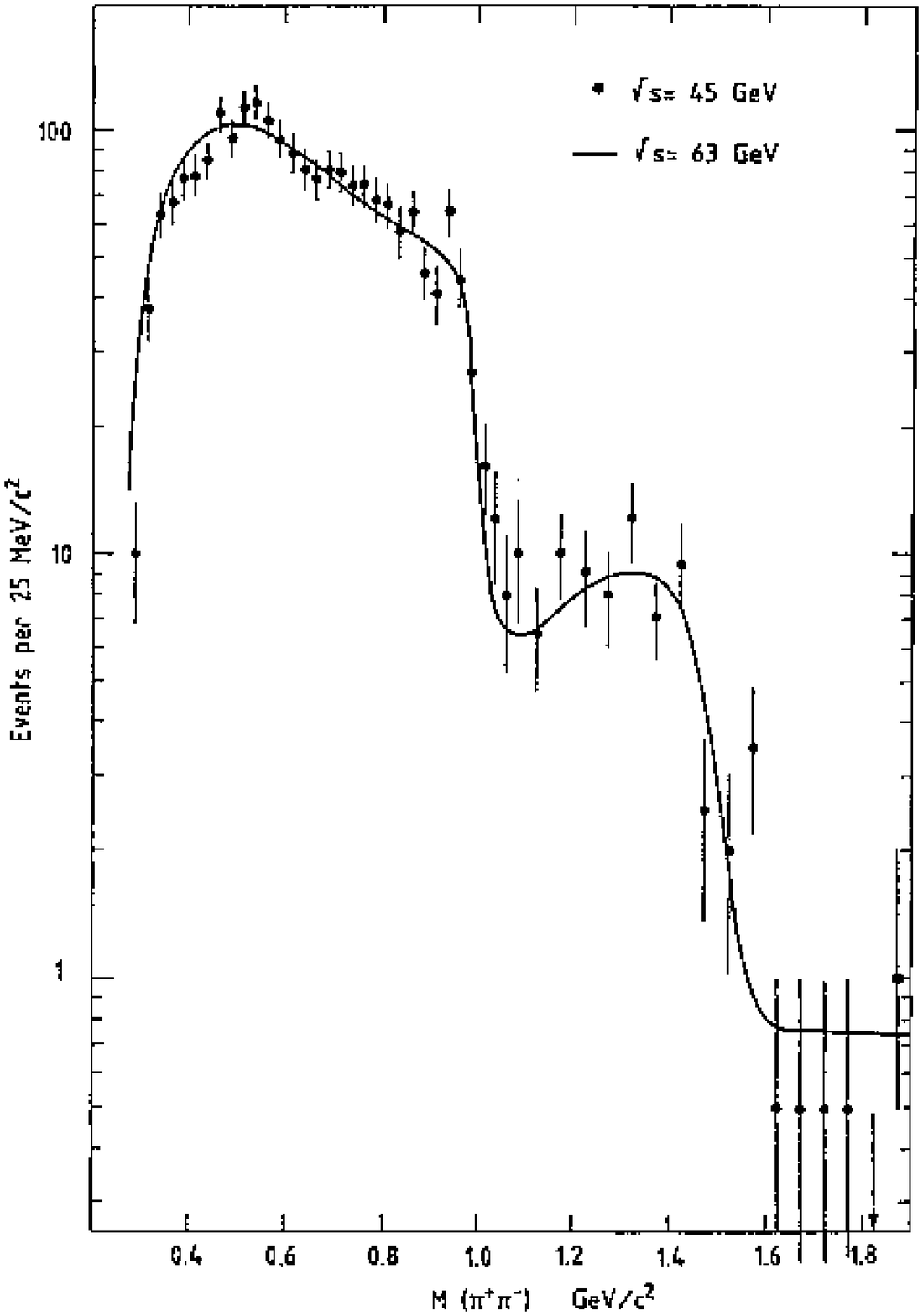,width=120mm}}
\end{center}
\caption{The mass spectrum  of the $\sqrt{s}$=45 GeV exclusive $\pi^+ \pi^-$
events (data points). The solid line represents the $\sqrt{s}$=63 GeV data,
normalised to the same total number of events. 
No acceptance has been applied to either distribution
\protect\cite{akesson,cecilphd}.
}
\label{figcent}
\end{figure}


\begin{figure}[htb]
\begin{center}
\vskip 1mm
\leavevmode
\resizebox{!}{16.0cm}{%
\includegraphics{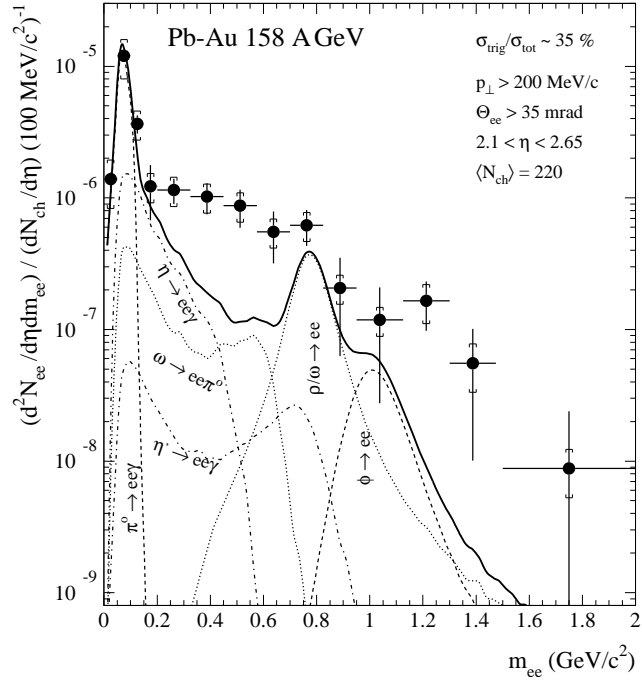}}
\vskip -2mm
\end{center}
{\center{
 \hspace*{1.0cm}
\begin{minipage}{13.cm}
\center{{\hspace*{-0.3cm} \mbox{} 
}} 
\begin{flushleft}
\caption{ 
Inclusive invariant $e^+ e^-$ mass spectrum
in 158 A GeV Pb+Au collisions normalised to the observed charge particle
density \protect\cite{na45}.
The statistical errors of the data are shown as bars,
the systematic errors are given independently as brackets.
The full line represents the 
$e^+ e^-$ yield from hadron decays scaled from p-induced collisions.
The contributions of individual decay channels are also shown.
}
\label{na45}
\end{flushleft}
\end{minipage} }}
\end{figure}


\begin{figure}[htb]
\begin{center}
\vskip 1mm
\leavevmode

\resizebox{!}{16.0cm}{%
\includegraphics{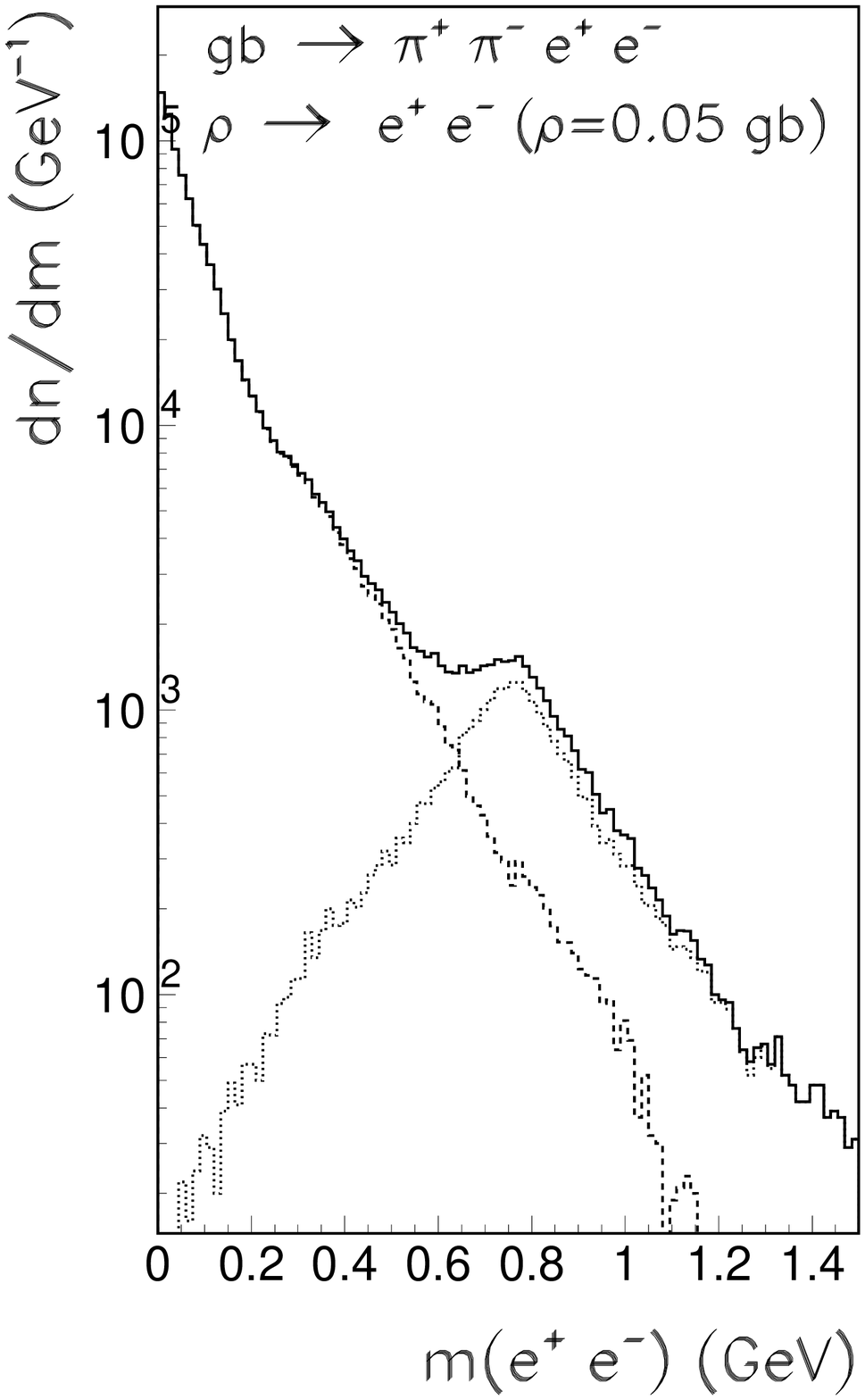}}
\vskip -2mm
\end{center}
{\center{
 \hspace*{1.0cm}
\begin{minipage}{13.cm}
\center{{\hspace*{-0.3cm} \mbox{ }
}} 
\begin{flushleft}
\caption{ 
Invariant mass distribution of $e^+ e^-$ pairs
resulting from the decay $\rho \rightarrow e^+ e^- $ 
and from the assumed decay of the $0^{++}$ glueball
  state $0^{++} \rightarrow \pi^+ \pi^- e^+ e^- $.
In this calculation
  the products of production cross section times
 branching fraction into
 $e^{+} e^{-}$ of the $\rho^{0}$ meson and the  $0^{++}$
glueball state are in the ratio of 1 : 20.
}
 \label{gb_rho_nocuts}
\end{flushleft}
\end{minipage} }}

\end{figure}


\begin{figure}[htb]
\begin{center}
\vskip 1mm
\leavevmode
\resizebox{!}{16.0cm}{%
\includegraphics{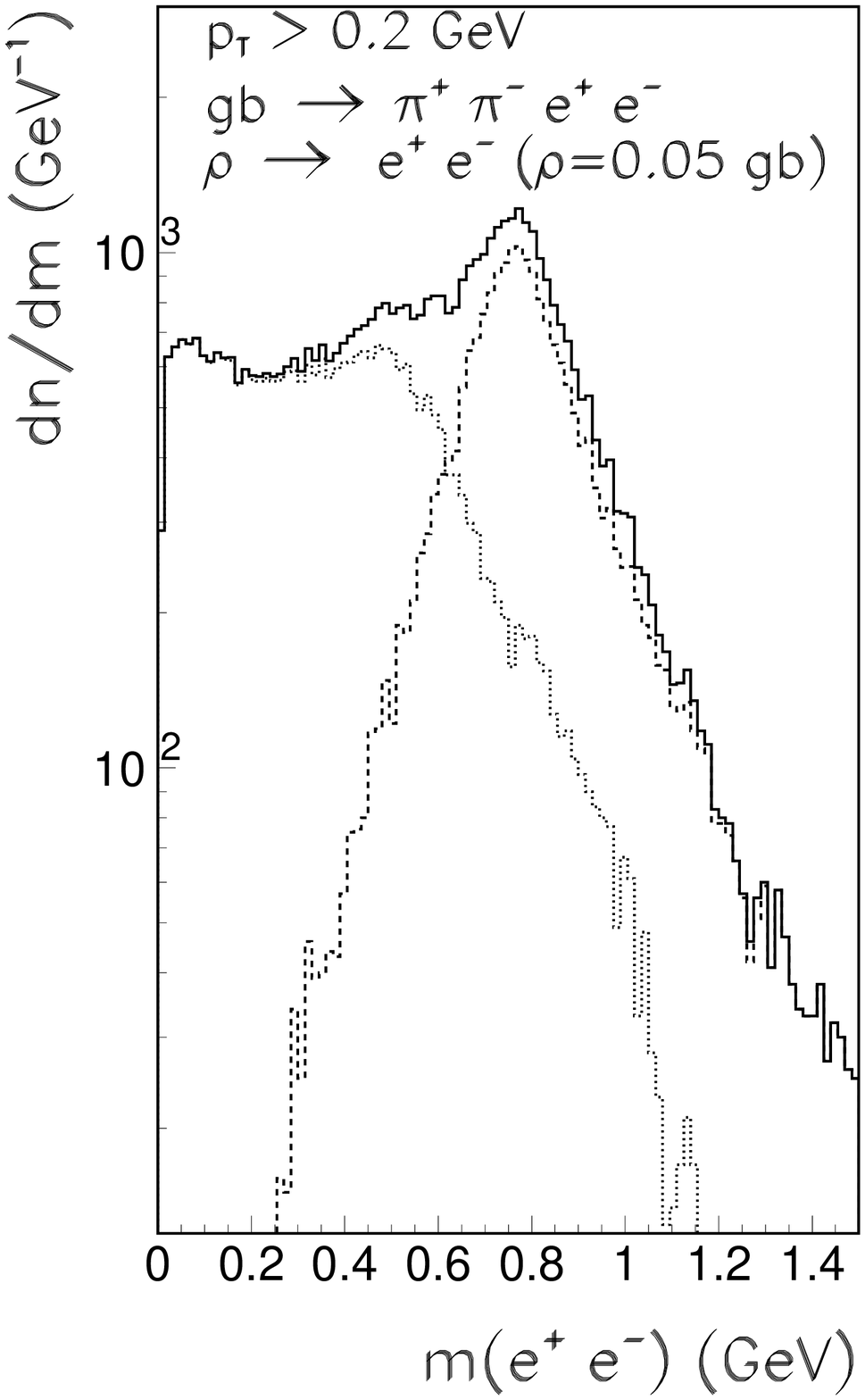}}
\vskip -2mm
\end{center}
{\center{
 \hspace*{1.0cm}
\begin{minipage}{13.cm}
\center{{\hspace*{-0.3cm} \mbox{ } 
}} 
\begin{flushleft}
\caption{ 
Invariant mass distribution of $e^+ e^-$ pairs
resulting from the decay $\rho \rightarrow e^+ e^- $ 
and from the assumed decay of the $0^{++}$ glueball
  state $0^{++} \rightarrow \pi^+ \pi^- e^+ e^- $.
  A cut on the transverse momentum of the diplepton pair
  of 0.2 GeV is imposed.
In this calculation
  the products of production cross section times
 branching fraction into
 $e^{+} e^{-}$ of the $\rho^{0}$ meson and the  $0^{++}$
glueball state are in the ratio of 1 : 20.
}
 \label{gb_rho_ptcut}
\end{flushleft}
\end{minipage} }}
\end{figure}


\begin{figure}[htb]
\begin{center}
\vskip 1mm
\leavevmode

\resizebox{!}{16.0cm}{%
\includegraphics{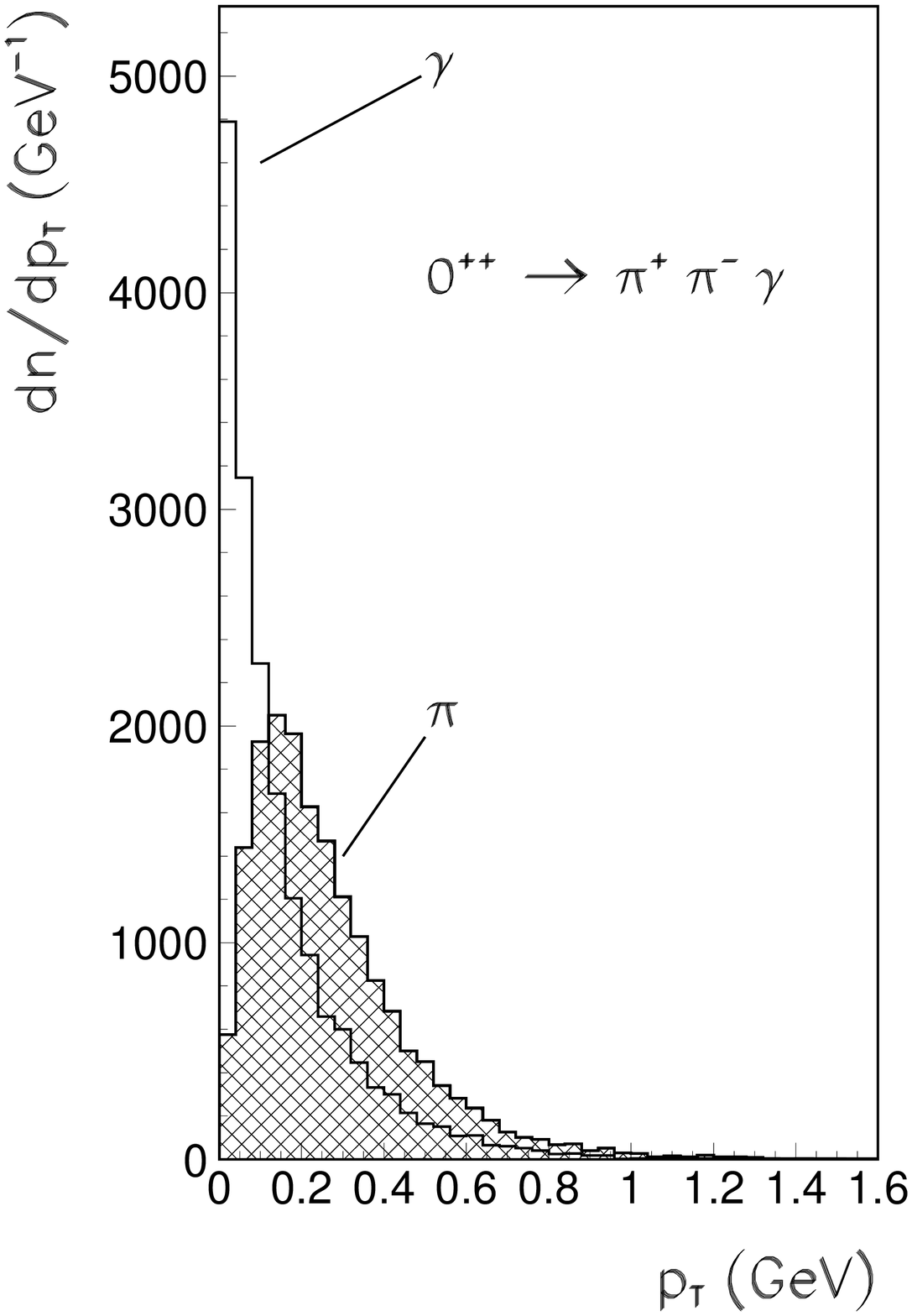}}

\vskip -2mm
\end{center}
{\center{
 \hspace*{1.0cm}
\begin{minipage}{13.0cm}
\center{{\hspace*{-0.3cm} \mbox{ }
}} 
\begin{flushleft}
\caption{
Transverse momentum distribution of photons and pions
 from the decay of the $0^{++}$ glueball state into
 $\pi^+ \pi^- \gamma $.
}
 \label{gb_to_gamma}
\end{flushleft}
\end{minipage} }}

\end{figure}


It is characterized by a different distribution over the two pion 
invariant mass than
the two to two scattering cross sections in figure \ref{gbfig3}. 
 The difference in shape is a manifestation of the
chameleon nature of the leading scalar glueball as the
unique genuinely wide resonance.
The destructive interference
pattern at the $q \ \overline{q}$ scalar resonances $f_{0} \ (980)$ and $f_{0} \ (1500)$ 
is clearly visible in both production line shapes. 

\end{description}

\noindent
The production of $gb \ ( \ 0^{++} \ )$ in the decays

\begin{description}

\item
c) $J/\Psi \ \rightarrow \ \gamma \ X \ ; \ X \ \rightarrow \ \pi \ \pi \ , \ K \ \overline{K}$

\item
d) $J/\Psi \ \rightarrow \ \omega \ X \ ; \ X \ \rightarrow \ \pi \ \pi \ , \ K \ \overline{K}$

\item
e) $J/\Psi \ \rightarrow \ \phi \ X \ ; \ X \ \rightarrow \ \pi \ \pi \ , \ K \ \overline{K}$

\end{description}

\noindent
where it was systematically searched for,
 remains unresolved for the time beeing.
\vspace*{0.2cm}

\noindent
The Axial Field Spectrometer collaboration
\cite{akesson} performs a spin analysis of the $\pi \pi$ system 
up to an invariant mass of 2.3 GeV, involving s and d waves. 
This analysis \cite{akess2} reveals a spin 2 resonance with the mass and width parameters 
$m \ = \ 1700 \ \mbox{MeV} \ , \ \Gamma \ = \ 120 \ \mbox{MeV}$. These parameters agree very well
with those of $gb \ ( \ 2^{++} \ )$ in table \ref{tabsum}.

\noindent
There is a controversy over the spin of the $f_J \ (1700)$ (J=0, 2)
 \cite{wa102_new,e690,akesson}.
 We refer to the review of particle physics
of the particle data group \cite{PDG}
and to ref. \cite{PMWO} for more details. Further experiments and/or analysis are
necessary to resolve this issue.

\section{Modes of production and experimental signatures of glueballs}

\noindent
We consider hadron hadron and nucleus nucleus 
collisions at  energies high enough to be dominated
by Pomeron exchange.
In this energy domain,
multiplicities of produced particles, including heavy flavors such as strangeness and charm,
could exhibit limiting behaviour in their ratios.
In this regime
we expect the glueball production to be dominant
without imposing rapidity gaps\footnote{Under rapidity gap we understand
a cut excluding events with any particle within characteristic intervals of
rapidity.}.

\noindent
We propose to study consequences of the
hypothesis that correspondingly 
dominant particle production proceeds from a gluon rich environment.
This then translates into a major component of $gb \ ( \ 0^{++} \ )$ production. 
In this process the lower mass region $m \ ( \ gb \ ( \ 0^{++} \ ) \ ) \ \sim 
400 \ \mbox{MeV}$ 
is assumed to be dominant, similar to the shape of the 'red dragon' observed in central
production in the reaction $p \ p \ \rightarrow \ p \ X \ p$ shown in figure
 \ref{figcent}.
It is a general property of a wide resonance that its line shape is not
universal but rather chameleon like.
This is supported by the contrast between the line shapes in
figures 1 and 2.
The line shape in figure 2 shows an exponential dependence on
the invariant mass, suggestive of a thermal distribution.
We therefore deduce that the production of low masses of the $0^{++}$
glueball state is preferred in a high energy and high density hadronic
environment.

\noindent
Thus the two pion decay mode of $ gb \ ( \ 0^{++} \ )$ becomes dominant, giving rise
to characteristic two pion s-wave correlations relative to directly produced pions, which 
in an approximately thermal and noninteracting model remain uncorrelated. 
The glueball associated s-wave component should dominate over direct
production of vector mesons $\varrho \ , 
\ \omega \ , \ \phi$ , with the main two pion decay
proceeding through the $\varrho$.

\noindent
The selection of central events, i.e. events 
with a small impact parameter 
enhances the gluon initiated production of $ gb \ ( \ 0^{++} \ )$ .

\noindent
We study the following production:
\begin{description}
\item 
 \vspace*{-0.6cm} 
\begin{equation}
\label{eq:1a}
\hspace*{-0.5cm}  
 \begin{array}{l}
 \vspace*{-0.5cm} \\
\hspace*{0.5cm}  \begin{array}[t]{lll}
\begin{array}{l}
h \ + \ h' 
\vspace*{0.3cm} \\
A \ + \ A' 
\end{array}
\ & \rightarrow &
\ N \ gb \ ( \ 0^{++} \ ) \ + \ n \ \pi \ + \ X
\end{array}
\end{array}
\end{equation}

 and the two decay channels:

\item 
 \vspace*{-0.6cm} 
\begin{equation}
\label{eq:2a}
\hspace*{-0.5cm}  
 \begin{array}{l}
1)
 \vspace*{-0.5cm} \\
\hspace*{0.5cm}  \begin{array}[t]{lll}
 gb \ ( \ 0^{++} \ ) \ \rightarrow \  \pi \ \pi,
  K \ \overline{K}
\end{array}
\end{array}
\end{equation}
\item 
 \vspace*{-0.6cm} 
\begin{equation}
\label{eq:3a}
\hspace*{-0.5cm}  
 \begin{array}{l}
2) \vspace*{-0.5cm} \\
\hspace*{0.5cm}  \begin{array}[t]{lll}
 gb \ ( \ 0^{++} \ ) \ \rightarrow \ \pi^{+} \pi^{-} \ \ell^{+} \ell^{-}
\end{array}
\end{array}
\end{equation}
\end{description}

\noindent
In eq. (\ref{eq:1a}) 
 N denotes  the number of scalar glueballs and n the number of pions
emitted  independently of glueballs.

\noindent
The semileptonic glueball decay in eq. (\ref{eq:3a}) proceeds as follows 

\begin{equation}
\label{eq:4a}
 \begin{array}{l}
 gb \ ( \ 0^{++} \ ) \ \rightarrow \ \varrho_{virtual} \ \gamma_{virtual}
\hspace{0.2cm}
,
\hspace{0.2cm}
 \varrho_{virtual} \ \rightarrow \ \ \pi^{+} \pi^{-} 
\hspace{0.2cm}
,
\hspace{0.2cm}
 \gamma_{virtual} \ \rightarrow \ \ell^{+} \ell^{-}
\end{array}
\end{equation}
  
\noindent
The decay channel in equation (\ref{eq:3a})
is not characteristic for a glueball.
The gluonic Zweig rule
suppresses the mixing of glueball and $q \overline{q}$ states,
but it does not apply to a low mass $0^{++}$ state because
of the strong effective coupling.
This is reflected by the large width ($\sim$ 1 GeV)
of the $0^{++}$ glueball \cite{HFPM,PMWO}.
We estimate that the branching fraction for the decay in eq. (\ref{eq:3a}) is of the 
order of $10^{-5} \ - \ 10^{-6}$.
This branching fraction is of the same order as $0^{++} \rightarrow 
\gamma \gamma$. The latter is extracted from central $e^+ e^-$ production
of hadrons \cite{penington}.

\noindent
The measurement of the $0^{++}$ glueball state in the
high track density environment in the final state of
heavy ion collisions could be pursued by searching for the
typical inteference pattern displayed in figures 
(\ref{gbfig3}) and (\ref{figcent})
in the 
invariant mass distributions
of $K^0_s K^0_s$, $K^+ K^-$, $\pi^0 \pi^0$ and $\pi^+ \pi^-$.
The decay channel to kaons is less affected by background, however
it can reconstruct only the tail of the $0^{++}$ invariant
mass distribution.
Especially the $K^0_s K^0_s$  mode singles out natural spin  parity
(e.g. $0^{++}$, $2^{++}$, $4^{++}$,...).
This fact compensates for the loss of statistics
as compared to the $K^+ K^-$ channel.
In particular the $J^{PC}=1^{--}$ channel is absent.
\\

The
 characteristic interference pattern could be enhanced applying 
cuts on kinematic variables like the transverse momentum ($p_T$)
and  the opening angle of the decay products.
Further we put forward the question, whether background subtraction
differential in relative momentum of the decay products (e.g. $K^0_s K^0_s$),
renders possible an angular momentum analysis overcoming the serious
combinatorial background.
This analysis has been succesfully performed in pp collisions 
\cite{akesson,cecilphd} albeit imposing severe rapidity gaps.
 \\

\noindent
The measurement of the $2^{++}$ glueball state through e.g. the decay channel
$2^{++} \rightarrow K \overline{K} $
may  be possible.
The associated production of the gb $2^{++}$
is suppressed compared to the $0^{++}$ state due to its mass.
Nevertheless it could be identified especially
in the $K \overline{K}$ channels because of two things:
first it would dominate the production of $2^{++}$ 
$q \overline{q}$ states of comparable mass and
second it would decay  through a normal decay width given in table 1 into
$\pi \pi$ and  $K \overline{K}$.
Due to the well defined shape of the  $2^{++}(1700)$  glueball 
and its small width  it is possible to suppress the
background in a more efficient way than for the $0^{++}$ glueball.
This is achieved through cuts applied to 
the decay kinematics of the meson pairs (e.g. $2^{++}(1700)  \rightarrow
 K \overline{K}$).
\\

\noindent
The production of the $0^{++}$ state could have important consequences for
the invariant mass distribution of $\ell^+ \ell^-$ pairs, through the 
decay channel $\pi^{+} \pi^{-} \ \ell^{+} \ell^{-}$.
The invariant mass distribution of $e^+ e^-$ pairs
produced in S+Au and Pb+Au reactions
at 200 and 158 GeV per nucleon respectively 
shows a significant excess over expectations based on
known sources of $e^+ e^-$ pair production \cite{na45} as
shown in figure \ref{na45}.
\\
   
\noindent
We emphasize that the contribution denoted 'known sources' in figure 
\ref{na45} represents the best knowledge as processed in Monte Carlo
studies of $p \ p$ and $p \ A$  collisions at comparatively lower energy density
than in central Pb Pb collisions at 158 GeV/c per nucleon.

\noindent
Figure \ref{gb_rho_nocuts} shows
the invariant mass distribution of $e^+ e^-$ pairs
resulting from the decay $\rho \rightarrow e^+ e^- $  
and from the assumed copious production and decay of the $0^{++}$ glueball 
state : $gb \ ( \ 0^{++} \ ) \rightarrow \pi^+ \pi^- e^+ e^- $.  \\

\noindent
In this calculation 
it was assumed that the product of production cross section times branching fraction
into $e^{+} e^{-}$ of the $\rho^{0}$ meson and the  $0^{++}$ glueball state 
are in the ratio of 1 : 20.

\noindent
The branching ratio of the $\rho \rightarrow e^+ e^-$ is $4.5 \ 10^{-5}$ and comparable
with  the one of $0^{++} \rightarrow \pi^+ \pi^- e^+ e^- $.

\noindent
Furthermore it was assumed that the  $0^{++}$ glueball state is distributed in mass 
as shown in figure  \ref{figcent}. 
The rapidity distribution is taken as gaussian with a width of 0.6 
units. For the transverse mass distribution we choose 
 an exponential shape with inverse slope of 150 MeV.

\noindent
The $\rho$ meson was assumed to have the same rapidity distribution as the
$0^{++}$ and its inverse slope in $m_{T}$ is taken as 200 MeV.
\noindent
Additionally in order to simulate the experimental response
we introduced an error of $\Delta p /p$ = 5$\%$  for the momenta
of the decay products $e^+ e^-$.

\noindent
In figure \ref{gb_rho_ptcut} the influence of the rejection of all
dilepton transverse momenta below 0.2 GeV is shown. 
This cut follows the actual procedure applied to the experimental
data of figure \ref{na45}.
The $e^+ e^-$ pairs from the decay of the $0^{++}$ glueball
state significantly populate the region of invariant mass where the excess is 
seen by the NA45 experiment.

\noindent
The decay channel $ 0^{++} \rightarrow \pi^{+} \pi^{-} \gamma$
occurs with a branching ratio typically 100 times larger than for the
decay in eq. (\ref{eq:3a}).
The resulting $\gamma$'s  would have predominantly very low transverse
momenta  displayed in figure \ref{gb_to_gamma}.
An abnormal production of $\gamma$'s from the above decay
is negligible compared to Bremsstrahlung and $\pi^0$ decay.
\\

\section{Conclusions}

\noindent
We discussed consequences of dominant   
 $ 0^{++}$ glueball production in
high energy collisions of hadrons and nuclei.

\noindent
It is conceivable that the  glueball
 production becomes a dominant part in central
nuclear collisions.
The existence of the QGP phase under thermodynamic conditions
above a transition temperature of 150-200 MeV
associates directly a gluon rich environment favoring the glueball production 
mechanism in the nuclear reaction case over
hadronic collisions at the same energy.

\noindent
The characteristic interference pattern of the 
$0^{++}$ glueball with the $f_{0} \ (980)$ and $f_{0} \ (1500)$
could serve as the signature of this state decaying into a pair of mesons 
 $K^0_s K^0_s$, $K^+ K^-$, $\pi^0 \pi^0$ and $\pi^+ \pi^-$
 without imposing large rapidity gaps.
However the two meson channels  suffer from combinatorial problems,
which may be reduced for the $K \overline{K}$ channels.

\noindent
A promising decay channel appears:
 $gb \ ( \ 0^{++} \ ) \ \rightarrow \ \pi^{+} \pi^{-} \ \ell^{+} \ell^{-} $.
This is not a characteristic glueball decay channel but enhanced through 
$q \overline{q}$ mixing
by the large width of glueball $0^{++}$.
Because the branching fraction of this channel is of the order $10^{-5} \ - \ 10^{-6}$
the opposite sign dilepton signal becomes significant compared to the $\rho^0$
contribution.
The expected signal  is dominant in the dilepton mass region between
200 and 800 MeV.
We conjecture that this enhancement is observed in S+Au and
Pb+Au collisions at 200 and 158 GeV  per nucleon respectively at low impact parameter
and high energy density.

\noindent
A dominant $0^{++}$ glueball component 
brings into focus an enhanced production of
glueball $2^{++}$(1700).
Due to the well defined shape of the $2^{++}$(1700) glueball and its 
small width it appears possible to suppress the background 
in a more efficient way than for the $0^{++}$ glueball.

\vspace{0.5cm}

\noindent
{\large \bf Acknowledgments}

\noindent
We would like to thank Ferenc Niedermeyer and Wolfgang Ochs
for discussions and the critical reading of the
manuscript.


\end{document}